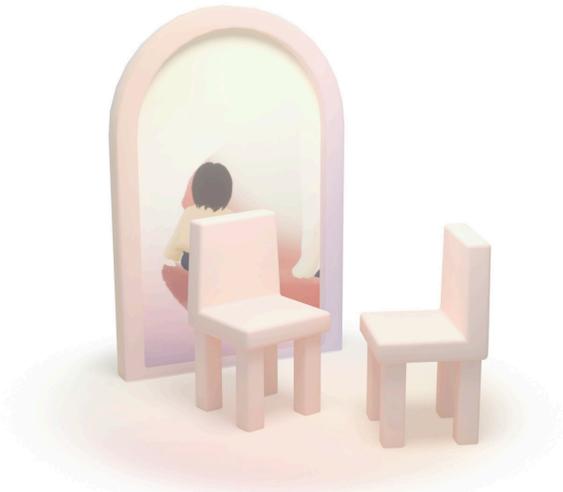

# MetaphorChat: A Metaphorical Chatting Space for Expressing and Understanding Inner Feelings

**Chen Ji**
University of California, Santa Cruz
jjjjjc12@gmail.com

**Katherine Isbister**
University of California, Santa Cruz
katherine.isbister@gmail.com

## ABSTRACT
Metaphors have been used during therapy sessions to facilitate the communication of inner feelings between clients and therapists. Can we create a digital metaphorical chatting space for daily use within close relationships? As the first step towards this vision, this work follows the autobiographical design approach to prototype *MetaphorChat*, which comprises two metaphorical chatting scenes tailored to meet researchers' genuine needs for discussing specific life topics in close relationships. Along with typing-based chatting, each scene offers a metaphorical narrative experience, composed of graphics and sound, with interactive mechanisms that deliver metaphorical meanings. This pictorial details the process of mapping abstract feelings into metaphor concepts, then how these concepts are translated into various interaction design elements, and the reflections from self-usage. We discuss the vision for such a metaphorical chatting space, uniquely positioned between messaging apps and video games, for the future design of empathetic communication applications.

## CSS Concepts
• Human-centered computing~Interaction Design

The New ACM 2012 Classifiers must be used:
https://dl.acm.org/ccs/ccs_flat.cfm

## INTRODUCTION
People want to feel understood when expressing their inner thoughts and feelings [23, 31, 40]. This need is particularly pronounced in intimate relationships, like those with family members and close friends, where feeling understood helps us seek support and build deeper connections [40]. Conversely, a persistent sense of not being understood can lead to alienation, potentially resulting in loneliness or worsened mental well-being over time [5, 22]. Feeling understood is important, but it is not a simple task. From the sharer's perspective, accurately articulating personal experiences and feelings is difficult, often leading to expressions of "fuzzy" feelings that cannot be effectively conveyed to our "support network." Additionally, from the receiver's perspective, family members from different generations and close friends from varied living environments, along with our unique personalities, can lead to diverse interpretations of the same experiences or feelings, hindering their understanding and support. This is likely the reason why, even though virtual communication has seamlessly woven into the fabric of our daily lives with minimal effort and great convenience, there are moments when we still find it difficult to express ourselves and feel understood [17].

Metaphors have been used as a communication tool in therapy sessions to facilitate the discussion and understanding of abstract life experiences and feelings between therapists and clients [2, 15, 34]. For example, a patient may describe her frustration using the metaphor "walking in a black hole" [15]. While frustration itself may be a vague concept, the metaphor helps therapists understand their clients by mentally simulating the experience it represents [10]. Beyond supporting understanding, metaphors can also offer transformative power. For instance, a therapist might offer "walking in a tunnel" as an alternative, acknowledging the current frustration stemming from the darkness, but also introducing the hope of an eventual exit from the tunnel [15]. Metaphors can be delivered not only in talking therapy but also through drawing, music, and sand tray therapy, employing different modalities such as visual, auditory, and embodied metaphors [35]. If multi-modal metaphors can aid in understanding and discussing abstract life experiences and feelings in the analog world, can we create a virtual space composed of digital metaphors to achieve the same goal?

Video games, as a medium, offer players a multi-sensory experience with metaphors conveyed through audio-visual elements and metaphorical game mechanics [8]. For example, *Journey* [37] invites players to experience the metaphor of *life as a journey* by having them walk alone in the sand, where they might randomly encounter another player. Additionally, the availability of in-game chatting allows players to not only experience these multi-sensory metaphors but also to share their feelings through chat bubbles. This raises a question: *Can we design a metaphorical chatting space, inspired by games, that incorporates audio-visual metaphors and interactive mechanisms, with a focus on person-to-person communication?*

Toward imagining and designing a metaphorical chatting space to express and understand inner feelings within close relationships, a profound understanding of one's own experience plays a critical role. Then, combining the roles of designer, researcher, and evaluator into one appears to be a good starting point, and the lessons learned can be expanded for more customizable design. Therefore, we followed the autobiographical design research approach [26], defined as *design research drawing on extensive, genuine usage by those creating or building the system*. This approach is beneficial for several reasons: First, having the researcher, designer, and evaluator roles filled by the same person ensures a deep understanding of genuine needs and the knowledge to translate these needs into design elements. Second, in intimate spaces, third-person methods' limitations might hinder deeper design insights. For example, participants might not fully disclose their genuine needs related to personal experiences, and building trust can be time-consuming. Third, this new design exploration benefits from the fast, iterative tinkering inherent in autobiographical design. To ensure academic rigor, criteria based on prior works [7, 25] were adhered to, as detailed in the Methodology.

This pictorial presents the researcher's journey in crafting two metaphorical chatting spaces *On a Boat* and *On a Train*: (1) identifying the genuine need to discuss certain complicated feelings; (2) detailing the process of deciding on metaphor mapping; (3) translating metaphor concepts into interaction design elements, such as text, graphics, sound, music, and interactive mechanisms; (4) reporting on the self-use process, evaluation, and reflection. Self-use feedback suggests this system as a slow and therapeutic space rather than a fast-paced messaging app. We discuss the vision for such a metaphorical chatting space, positioning it between messaging apps and video games as a unique medium for empathetic communication. Finally, we summarize our design approach on how to integrate metaphors into a metaphorical chatting space.

We view this work as an initial step in a broader investigation of using metaphor as a tool for empathetic and therapeutic communication design. We hope our documentation of crafting *MetaphorChat* will offer both conceptual and practical inspiration.

## RELATED WORKS
### Autobiographical Design
While mainstream third-person design methods in HCI, such as participatory design, lab or field evaluation, focus on objective, third-party knowledge, Neustaedter and Sengers [26] argue that designing a system with oneself as the target user and evaluating it through personal usage can offer valuable lessons. They refer to this approach as *autobiographical design*. Researchers have applied this method to create personalized, rich experiences, including designing for domestic use (e.g., *Video Window* [9], *Leaky Objects* [11]) and in designs within intimate relationships [4, 12].

To define the methodology and foster academic rigor, Neustaedter, Sengers, and Judge [25] have formalized five tenets of this method: *Genuine Need* (requiring authentic motivations), *Real System* (the system must be developed and functional), *Fast Tinkering* (enabling designers to quickly test ideas, iterate, and refine), *Record Keeping and Data Collection* (documenting the detailed design process provides insights for design research and assists other researchers), and *Long-term Usage* (suitable for long-term research, which can lead to a more holistic understanding of a design's impact). Desjardins and Ball [7] provide further critical insights and suggestions for using autobiographical design:

- *Genuine Needs*: Personal interests, curiosity, imagination, and research goals are important in initiating autobiographical design projects. For instance, Gaver's motivation for his *Window project* [8] stemmed from a blend of curiosity, technological potential, and aesthetic creativity.
- *Design Participation*: Design participants often include the main researcher's network, like family and friends as co-designers, and research team members as observers. There's a lack of examples where a designer is designing for themselves, which might offer unique, introspective research insights.
- *Intimacy*: Autobiographical design often delves into personal or taboo topics, necessitating clear boundaries on shared content. It is beneficial to focus the audience's attention on design insights rather than on personal and emotional writing.
- *Reflexivity*: It is crucial to record the process of why certain design decisions were made for reflexivity

### Metaphor for Supporting Communications
Metaphors have been applied in HCI for supporting communication due to their descriptive and generative nature. Reed et al. [30] demonstrated that abstract metaphors can effectively convey experiences and convey information, as illustrated by vocal teachers who use metaphors to teach singing. Metaphorical representations of emotional states in emotional regulation and mindfulness designs, though not explicitly mentioned, can be seen as forms of self-talk about inner feelings [6, 21, 32]. For example, [21] employs the degree of darkness and roughness of the floor as metaphors for anxiety. Additionally, some self-reflection applications use metaphorical elements as expressive tools for user introspection, exemplified by weather representing mood [38], anger as fire [39], personal growth as a tree [39], and life as paper [1]. These works are more intentionally designed for self-talk about inner feelings. The creation of a digital metaphorical space for communicating inner feelings within close relationships remains under-explored.

### In-Game Social Communictaion
Games research [14, 16] has suggested that players often disclose personal information and create connections in games, leading to online strangers frequently becoming offline friends due to in-game communication [18]. For example, [41] analyzed players' use of *Animal Crossing: New Horizons* [27] during COVID-19, highlighting how it satisfies basic psychological needs such as free expression and fosters the building of new connections through chatting and engaging in activities together, like watching fireworks or decorating neighborhoods. The social aspects of *Sky: Children of the Light* [36] resemble real-life friendship mechanisms: chatting, dancing, and exploring together. Additionally, virtual spaces, like park benches or picnic tables, allow people to more easily share common spaces and initiate conversations [3].

We see the potential of metaphors in supporting emotional communication through technology and view video games as a medium for both chatting and experiencing multi-sensory metaphors. Therefore, our work aims to design a metaphorical chatting space that features game elements, including metaphorical audio-visual elements and interactive mechanisms, while focusing on communication rather than gameplay.

# Methdology

We applied an autobiographical design approach [26], taking into account the five tenets of autobiographical design from Neustaedter, Sengers, and Judge [25] and critical insights on how to address nuanced consideration from Desjardins and Ball [7]. We evaluated our design motivation, design participation, and the considerations in keeping our design process, as well as our presentation language.

## Genuine Need

The motivation for designing this digital metaphorical chatting space is as follows:

✦ **Curiosity**
Metaphors have been used in therapy sessions for communicating inner feelings. What if metaphors were integrated into daily chatting applications for close relationships?

✦ **Imagination**
I imagine a metaphorical chatting space that delivers metaphors through various design elements such as graphics, sound, music, and interactive mechanisms for discussing inner feelings.

✦ **Reserach Goal**
○ Gaining insights on how we can integrate metaphors for supporting expressing and understanding inner feelings.
○ Proof of concept of a metaphorical chatting application in the space between message apps and video games.

## Real System

It is a functional system, which has been used between the first anonymous author's close friends. Since the system is envisioned as a chatting space with chatting functionality and interactive audio-visual experience similar to video game, Unity was used to implement these chatting spaces. Photon 2 was used to simulate the realtime chatting functionality, and the synchronization of graphics and interaction in the scene. The visual assets were drawn in Procreate, ambient sound was downloaded from freesound, and the music was composed in Logic Pro X.

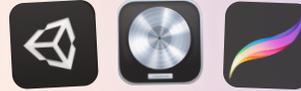

## Documenting

Documentation is crucial in the autobiographical design approach, serving both self-reflection and the potential to inspire other researchers. The main researcher maintained a design diary using Figma, documenting and reflecting on ideas, life experiences that inspired the design, sketches drawn, possible UX layouts, and technical solutions used for implementation. During the self-use stage, the chatting experience was screen-recorded.

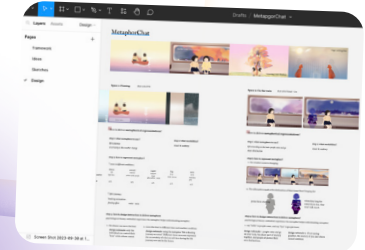

## Language

Discussing inner feelings inevitably includes some emotional language. Considering *intimacy* and *privacy*, we avoid specific event descriptions; Instead, we focus on more abstract or metaphorical language to depict these feelings or experiences. This approach can also serve as a showcase for using verbal metaphors in discussing inner feelings. In addition, we will concentrate on sharing the design process. This will include an explanation of why certain metaphors were chosen, based on the first author's experience, as well as outlining the design trajectories from concept to implementation.

## Design participation

The first author is the primary researcher, designer, and evaluator of the system, and their close friends act as co-users. The second author serves as an advisor, providing suggestions during the design process, and as an observer, assisting in the reflection on research insights.

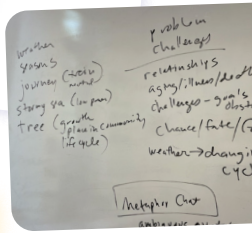

## Check List

- ✓ Genuine Need
- ✓ Real System
- ✓ Record Keeping
- ✓ Fast Tinkering
- ✓ Design Participation
- ✓ Intimacy
- ✓ Reflexivity
- ⊖ Long-term Use

This pictorial only records the first stage of using such metaphorical space. Long-term use will be reported in the future.

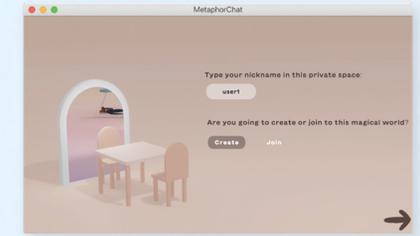
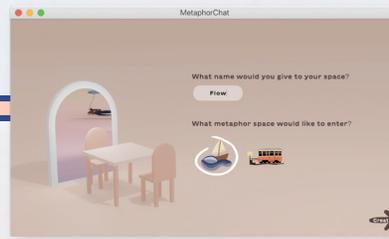
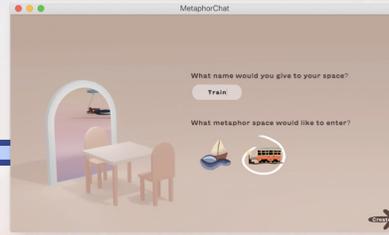
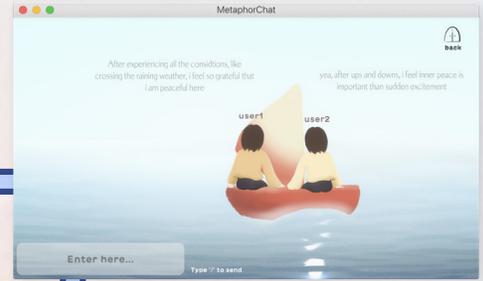
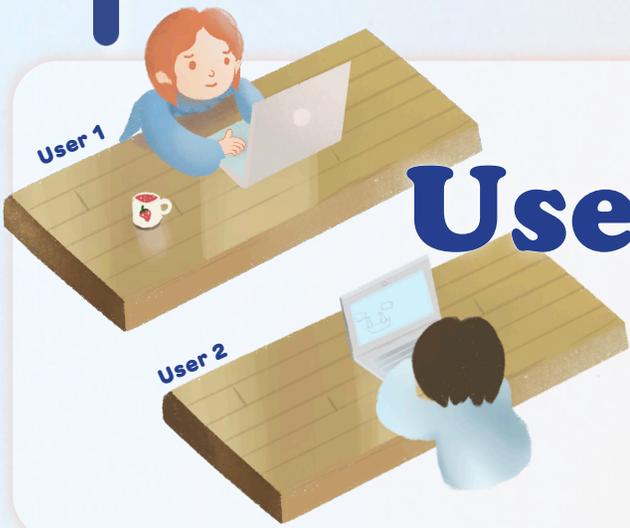

User 1 types user-name and creates a chatting space

User 1 types the room name, selects a metaphorical scene, and creates the room

User 1 and user 2 gathers in the chatting space *On a Boat*

# User Flow

This is a chat space where one user creates a chat room, and another joins. Here, we illustrate the user flow of how two users enter two metaphorical chat experiences, *On a Boat* and *On a Train*.

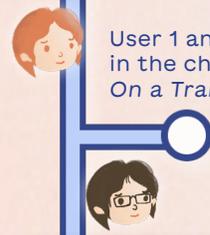
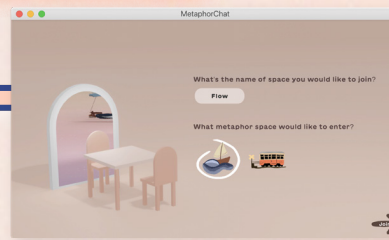
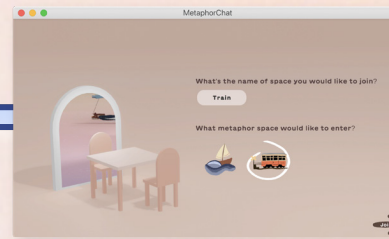
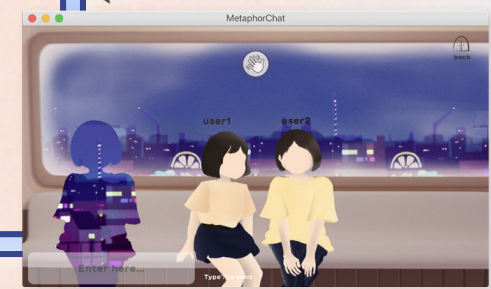

User 2 types user-name and joins a chatting space

User 2 types the room name, selects a metaphorical scene and joins the room

User 1 and user 2 gathers in the chatting space *On a Train*

# On a boat
(Design Diary)

## Needs
I find it difficult to discuss my feelings of **uncertainty and insecurities while pursuing my goals** with my "support network". I often resort to figurative language, such as feeling as lost as on a foggy day, to describe my emotions. However, words sometimes fail to capture the nuanced feelings I wish to convey. My parents and friends also seem to struggle with understanding these feelings of uncertainty and insecurity, which can lead to frustration in conversations that are meant to be supportive. Due to these barriers in expressing and receiving abstract feelings, I am motivated to create a metaphorical space for chatting that supports discussions about uncertainty and insecurities.

## Metaphor Mapping
When I want to express feelings of uncertainty, being lost, and sometimes hopelessness, I first thought of **life as a journey**, with **uncertainty akin to changing weather**. The metaphor of *life as a journey* can apply to many contexts, such as traveling on the ocean, on the highway, by train, or walking in a field. While I was ideating, I played a game called *Old Man's Journey* [33]. In one level, I navigated a boat on the ocean during a stormy night. This experience mirrored the sense of uncertainty and feeling lost that I wanted to convey to my family members. Therefore, I narrowed down the general metaphor of **life as a journey** to the specific experience of **floating on the ocean**.

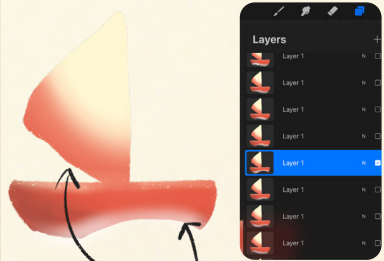

Additionally, a one-time kayaking experience with a friend made me realize that floating in the same boat could be a meaningful experience to design into my metaphorical space. **Inviting someone else into my boat** can serve as a metaphor for **inviting them to "feel" my feelings**.

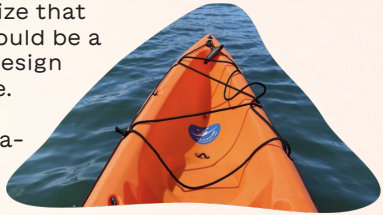

Then, I finalized the metaphor mapping as below:
**pursuing my goals** <--> **floating on the ocean (m1)**
**uncertainty** <--> **changing weather (m2)**
**inviting someone to "feel" my feelings**
<--> **inviting someone on my boat (m3)**

## How to translate these metaphors?

### Step 1: Explore visual Metaphor
I began by sketching some graphics in Procreate, which also included iterations to achieve my aesthetic goals.

**m1: floating on the ocean**   **m2: changing weather**

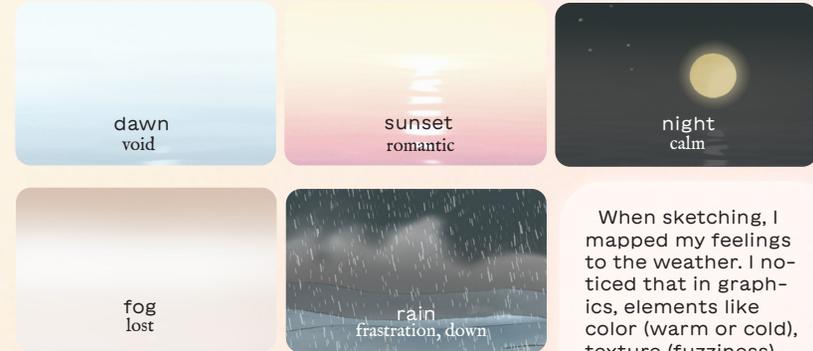

made *wind* and *wave* animation effect on boat to make it it feel like floating

dawn — void
sunset — romantic
night — calm
fog — lost
rain — frustration, down

When sketching, I mapped my feelings to the weather. I noticed that in graphics, elements like color (warm or cold), texture (fuzziness), and light can help me convey visual metaphors effectively.

### Step 2: Explore audio metaphor & interactive mechanism
Once I was satisfied with how static graphics looked, I exported them as assets and imported them into Unity to add sound and interaction.

**m3: inviting someone on my boat**

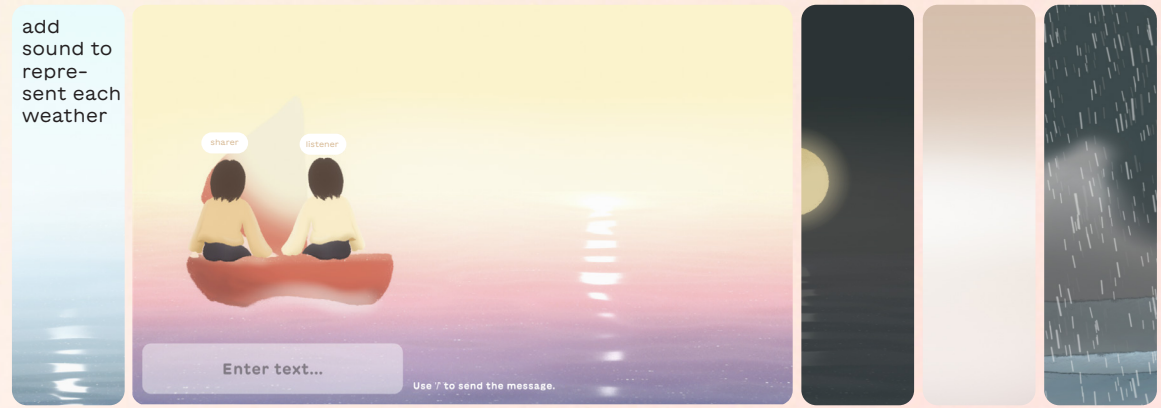

add sound to represent each weather

**Interaction 1: moving across different weather condition**
Inviting someone to join my boat as it crosses various weather conditions serves as a metaphor for inviting them to "feel" my feelings on my journey. Designed for a desktop experience, I used the "left arrow" and "right arrow" keys for input.

**Interaction 2: only sharer can move the boat**
My initial idea was to design the interaction so that both the sharer and the listener could control the boat across different weather conditions. However, I reflected on the fact that while people around me can provide support and advice, only I can make decisions or change my mind. Therefore, I modified the interaction so that only the sharer can move the boat, while the listener can only experience the transformation passively, without controlling the boat.

# On a boat

(Self-use)

## Self-use procedure

The main researcher invited her close friend, LJ, to try the **On a Boat** chatting experience. They used the system in a co-located setting, where both users operated their laptops and wore headphones to listen to the auditory elements. The main researcher, serving as the sharer, expressed her feelings of uncertainty and insecurities, while LJ acted as the listener. During the chat, they exchanged life experiences, which were mirrored by the changing weather and each other's words. At the end of the self-use session, the main researcher conducted a semi-structured interview to understand their feelings about the chatting experience and their views on such a metaphorical chatting space.

## Sharer's perspective

✦ **A way to invite the other person to experience my inner world, making it easier to initiate an emotional conversation**

Seeing my friend on "my boat" and traveling with me through "my mood weather" makes me feel as if they are experiencing my inner world - visually, auditorily, and in an embodied manner. As they accompany me, I can type more detailed experiences or feelings (e.g., feeling like navigating a boat in fog when trying to find my career goal, just like now), based on what they observe. This environment eases the initiation of emotional conversations by setting the stage for specific emotions, eliminating the need for figurative language to describe my feelings. Moreover, since we experience the same environment together, our conversation can expand based on the shared objects in the scene. These are more tangible and concrete compared to a scenario where we each imagine a scene without a common reference point. The metaphorical environment thus acts as a connecting bridge between our mental imagery, which not only supports the expression of my feelings but also makes it easier for the listener to understand, based on my perception.

✦ **Metaphors evoke some mindset discussion**

Being in the metaphorical scene enables us to unfold some unexpectedly deep discussions about how we face various situations. For example, while traveling in "rainy" weather, we talked about how "sometimes we cannot control bad weather." After traversing through severe weather conditions and returning to "dawn," we chatted about the realization that being peaceful and "bored" is actually a blessing.

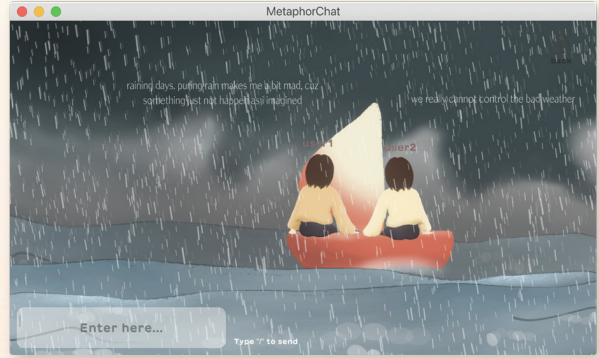

✦ **Emphasize the feeling of being together**

Visually seeing my friend beside me on the boat truly emphasized the feeling of being accompanied and listened to. Compared to the verbal metaphor "I am on the same boat as you," this interactive, multimodal experience of "being on the same boat" across my "emotional weathers" further enhances my sense of "togetherness."

## Listener's perspective

✦ **Virtual environment evoked memory supports understanding feelings from sharers**

LJ felt that changing the space acted as a callback for her; it recalled memories, connecting the weather and vibes with her previous experiences, which helped her correlate her past experiences to understand my feelings. 

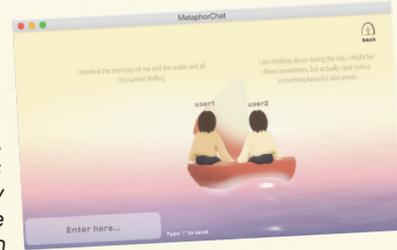

She said, "*A lot of things randomly pop up in my mind [...] I experience emotions associated with different colors [...] for instance, the pink one evokes the feeling of sunshine during the golden hour, bringing back memories of happiness.*" Such an experience is like introducing a unique environmental aspect to the scenario, which stimulates the listener's brain in the same way and enhances the shared imagination with the sharer and their expressions. Additionally, LJ also mentioned that transitioning through different weathers in just a minute is impossible in the real world. Here, however, we can experience such transitions in a short period, which is akin to a condensed representation of a life snapshot.

## Reflecting on Metaphor Theories

I reflected on why we felt this way, drawing on metaphor theories from linguistics and psychology [15, 19, 20, 35].

✦ **Metaphor reflects thoughts**

According to *Metaphors We Live By*, metaphors are not just a matter of language but also of thought [19, 20]. Language can be viewed as the embodiment of our thoughts; what we represent in our language reflects the structure of our thoughts. This is why metaphors have been applied in psychotherapy [15], where therapists use them to help people navigate their feelings, thoughts, and memories. In our case, both as a designer and a user, the process of creating such a metaphorical chatting space - from identifying metaphorical mappings of feelings to expressing and interpreting them into audio-visual metaphors and interactive mechanisms - is the process of me processing my feelings and thoughts via metaphors, similar to self-reflection. Then, I showcase my embodied feelings to my "supportive network".

✦ **Metaphors support understanding by converting abstract concepts to concrete ones**

Since metaphors can concretize abstract concepts [19, 20], they enable me to avoid complicated descriptions of abstract feelings and instead use concrete ones, such as a boat journey or weather changes. Therefore, this approach makes it easier for me to express myself as a sharer and also simplifies understanding for the listener.

Additionally, in the case of a metaphorical chatting space, audio-visual metaphors and interactive mechanisms seem to serve as a more straightforward format compared to using shared metaphorical language as an anchor to discuss feelings between users.

After prototyping **On a Boat**, I established my prototyping workflow as follows: 1. Metaphor mapping through collecting inspiration from life experiences, literature, films, or video games; 2. Sketching the graphics based on the metaphor; 3. Transporting the graphics into Unity to add interaction, sound, and music. Within Unity, I conducted fast iterations based on whether the interaction felt right or not, and whether it could effectively convey my inner feelings from my perspective.

# On a train
(Design Diary)

## How to translate these metaphors?
- Step 1: Explroe visual Metaphor
- Step 2: Explore audio metaphor & interactive mechanism

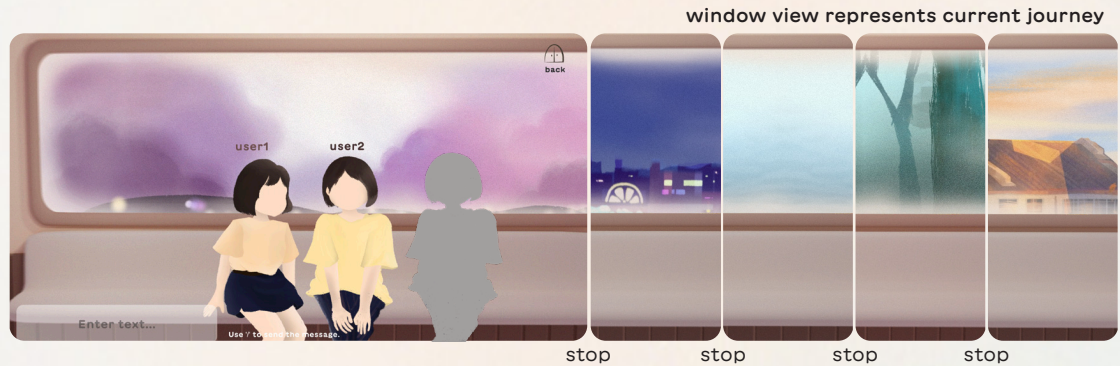

window view represents current journey

stop   stop   stop   stop

### Interactive 1: pressing the silhouette to chat
silhouette represents people come and go

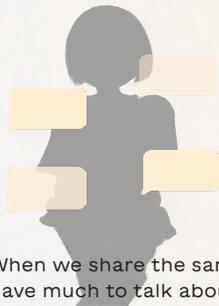

The interaction was inspired by the game *Florence* [24], where the designers convey arguments through fast-moving and red conversation bubbles.

When we share the same journey, we have much to talk about. In this interaction, pressing the silhouette each time will generate a chat bubble.

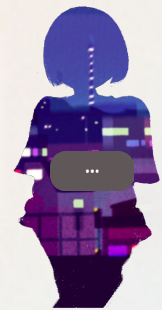

The silhouette's mask symbolizes the destination their heart truly desires. When people are exploring, they might not have a specific place in mind, which is why I designed it to be empty-grey upon their arrival (on the left). However, when they find their destination (e.g., a city), I design their current destination as a mask. The interaction here includes: (1) their mask transforms into a concrete destination; (2) the train stops; (3) when pressed, they no longer respond.

### Interactive 2: the train will move until you say bye

People come and go on the train. We share a part of the journey together, and everyone pursues their own destinations. If we don't say goodbye, neither our journey nor that of the others can continue. Through the design of an interaction, I hope to convey this meaning, where pressing the "bye" button allows the journey to continue.

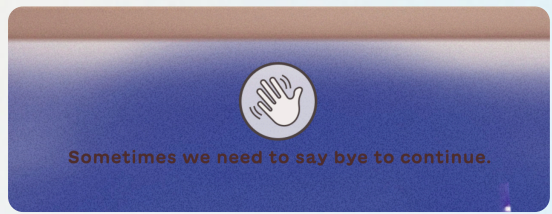

## Needs
Interpersonal relationships are common topics in the life experiences I usually discuss with my close friends. The concept of **coming and going** is a frequent and inevitable theme, often accompanied by feelings of frustration. Sometimes, when discussing such feelings, we use the metaphor, "**we are on our life train, and everyone has their destinations**." However, the discussion might not continue beyond that. Considering that something more nuanced could be expressed using graphics, sound, or interaction, I hope to create a metaphorical space for my close friends and me to discuss such topics, aiming to extend the conversation and also instill a positive mindset.

## Metaphor Mapping
The metaphor of **a relationship as traveling on the same vehicle** is commonly used to describe interpersonal connections. For instance, Lakoff [18] discusses the concept of a vehicle in his explanation of the metaphor **love as a journey**. He describes people in a relationship as travelers on a journey together, with their common life goals as their shared destinations. The relationship serves as their vehicle, enabling them to pursue those goals together. However, the journey isn't easy - there are obstacles, and decisions must be made about which direction to take and whether to continue traveling together. This metaphor applies to any type of interpersonal relationship.

When deciding which vehicle was more appropriate, I considered different options. A car seemed too private and not well-suited for all types of interpersonal relationships. A train, on the other hand, seemed like a good choice: it encounters many stops, where we might potentially meet random people, or some might decide to get off. The finalized metaphor mapping is shown below:

*relationship* <--> *travel on the train (m1)*

*people come and go* <--> *train's stops (m2)*

# On a train
(Self-use)

## Self-use procedure

The main researcher invited her long-time friend, TX, to try the **Train** chatting experience. They used the system in a remote setting, where they were located in different cities and logged into the system synchronously. The main researcher, serving as the sharer, expressed her feelings of people coming and going along the life journey, while TX acted as the listener, who also shared his thoughts on his feelings. At the end of the self-use session, the main researcher conducted a semi-structured interview to understand their feelings about the chatting experience and their views on such a metaphorical chatting space.

## Sharer's perspective

✦ **Experiencing the train journey together evokes discussion**

As alumni of the same high school, we discussed our attempts to re-establish connections with former classmates. However, these efforts proved futile as conversations rarely progressed beyond initial pleasantries. This outcome was unexpected for both of us, considering our perceived connections with others, which had been facilitated through interactions such as "comments" and "likes" on social media platforms. In reality, these connections were superficial. The conversation was enriching not only because I articulated my emotions about the transient nature of relationships using a train metaphor, but also because my friend shared similar experiences, fostering a sense of mutual understanding in this metaphorical space.

✦ **Window view as a trigger for destination sharing**

Although the intention was to discuss my complicated feelings about people coming and going along our journey, our conversation took a different turn. As we watched the passing window view while chatting, we encountered questions like, *"Do you know where you are heading?"* This led us to discuss our current destinations and past experiences of trying to live in different environments. For example, as the train passed through the city, I shared my thoughts: Although I was attracted by all the exciting opportunities happening in the city, I felt a longing for a place that calms me down and makes me feel grounded. TX also talked about his goal of being a free-walking person. The discussion revealed how the embedded journey metaphor in the window view could trigger reflections on our ideal destinations.

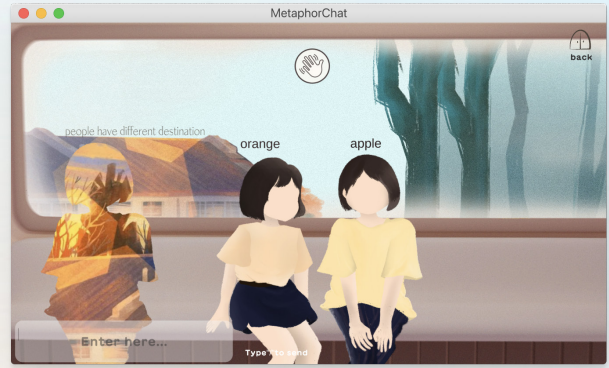

## Listener's perspective

✦ **The interactive mechanism helps understanding feelings from sharers**

TX mentioned that when I shared my feelings of "people having their own destinations in their hearts, and although we are physically together, I feel we are apart," he found it to be a really abstract expression. While he understood the general meaning, it still seemed vague to him. However, during the experience, he noted that when he saw the silhouettes' masks change and they stopped speaking to me with a "speechless bubble," he felt exactly what I was talking about *"this experience is like visualization of your feelings"*. He then said that he also resonates with my experience of people coming and going in my life's journey. It seems that the interaction can convey metaphorical meanings.

✦ **A private virtual simulation of a train ride**

While discussing the question, *"How about we chat about today's topic on a real train?"* TX said that it might be disturbing due to noise and crowds. He mentioned that having too many people on the train can make it feel less private, which could influence his motivation to express inner feelings and thoughts. This metaphorical space, while still featuring the train scene, offers privacy and provides a nice setting—with pleasant visuals and gentle music—for discussing life topics.

## Reflecting on Metaphor Theories

✦ **Metaphor reflects thoughts**

✦ **Metaphors support understanding by converting abstract concepts to concrete ones**

When experiencing the Train, TX and I talked about the window views, silhouettes, and our destinations. In reality, we were sharing our thoughts on life, having and losing, as well as re-navigating ourselves. These concrete representations make abstract topics become tangible.

✦ **Metaphor is understood by Experience**

As [10] said, we understand metaphors by comparing them with our own experiences. Here, the interaction design of pressing the silhouette to chat, and pressing the "bye" button to continue the journey, seems to provide a more engaging comparison with our past experiences. This might lead to a better understanding of the metaphor *life is a journey* and evoke a deep discussion on such topic.

## Usage Difficulty

Since the scenario here involves remote chatting, we've noticed that network issues sometimes occur, such as unstable connection states. In searching for the reason why the connection is not always successful, the PUN2 web pages suggest that connections tend to be smoother when users are in locations sharing the same server. Due to the unstable connection, the delivery of chatting content is sometimes delayed. Addressing this network issue in the future is necessary for a smoother experience.

## Envision: An slow space for emotional and theraputic conversation

During the interview after the self-use session, participants were asked how they felt about the differences between using messaging apps and such a metaphorical chatting space.

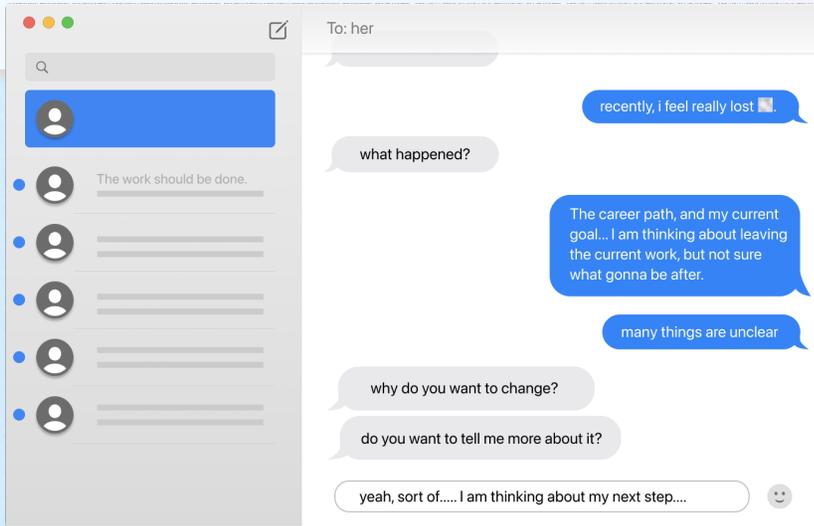 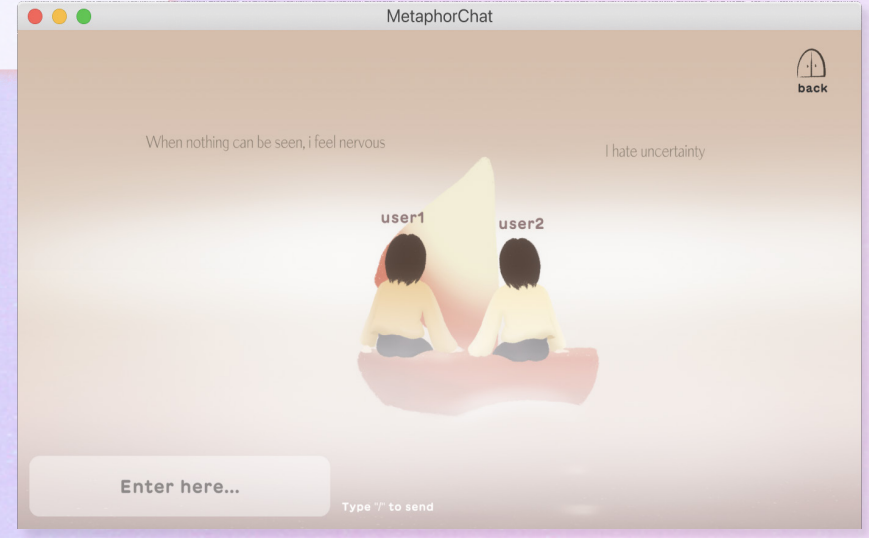

Social app is fast, keep the chat history sometimes feel overwhelming.

For productivity purpose.

Not for productivity

Sometimes, in social apps, it's expected to be specific when expressing emotions, as random emotional expressions may not always be well-received.

Slower than chatting app we used. Since the text bubble will disapear, i need to focus, which makes me slow down.

Can be emotional here, It is a space for healing.

I prefer not to express deep thoughts or feelings on social apps. Instead, I prefer to ask people out for coffee, a walk, or in a special analog setting in the real world.

It is like you entered a tunnel, you enter a space. My brain chills, with a lot of thoughts and memory randomly popping up.

## Toward Slow and Healing

As participants described it as a healing space, we explored how metaphor might play a role in making such experiences healing. Research on metaphors in psychotherapy [15] suggests that seeking mental healing is similar to physical healing, with the first step being the identification of the problem. Metaphors then become a useful tool for people to translate their complicated feelings into something more concrete, aiding in self-understanding. This process is completed during the design phase, and inviting friends to experience it is a way to revisit these feelings. Furthermore, metaphors [2, 15] can provide an indirect way for us to navigate through our feelings and thoughts. For example, we might say "saying goodbye on our train" instead of directly mentioning traumatic memories of a specific interpersonal relationship breakup. We can also bring therapeutic metaphors such as "it is not always bad weather" for constructing positive mindsets.

Additionally, aligned with the design agenda mentioned in *Slow Technology*, which emphasizes designing for slowness, mental rest, and a less consumptive lifestyle [28, 29], such a metaphorical chatting space is described as slower-paced, allows users to be more vulnerable, and is less focused on productivity. We summarized some slow features based on participants' feedback: the sound of water, the absence of typing recommendations, the disappearance of chat bubbles, and the muted aesthetics, which can be employed in the future design of slow communication tools.

## Envision: An In-between slow and healing space between message apps and video games

We also envision *MetaphorChat* as a space that bridges messaging apps and video games. It features chat functionality, multimedia metaphor representation (including graphics, sound, and music), and an interactive aspect.

I feel like climbing the mountain in a foggy day. It is white.

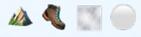

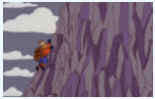

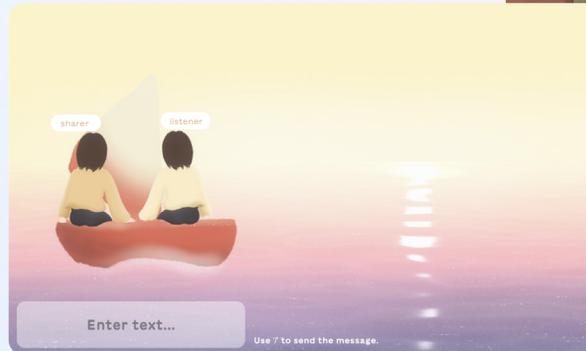

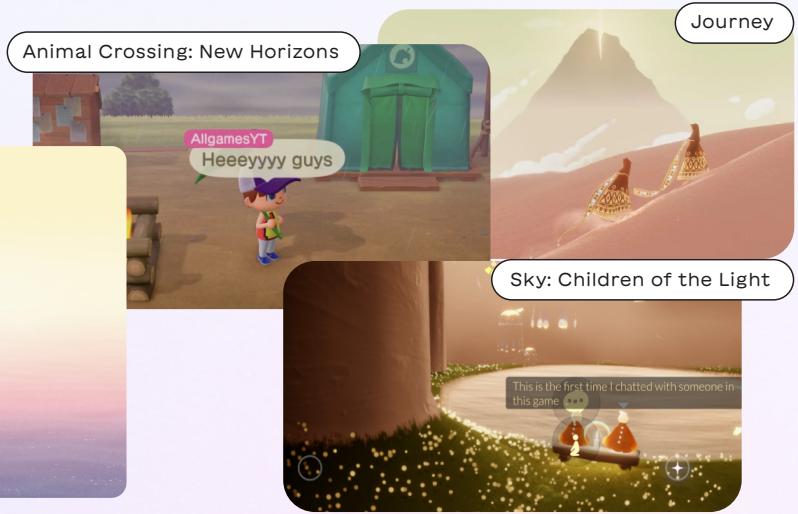

Animal Crossing: New Horizons

Journey

Sky: Children of the Light

**Message Apps**

**In-between**

**Video Games**

We compare three different mediums for delivering metaphorical communications in messaging apps and video games, as well as the in-between medium that we have crafted.

Let's consider a situation where we use current messaging apps to express our feelings through metaphors. For example, to convey the feeling of being "ungrounded," we might use metaphorical text such as, "I feel like I'm climbing a mountain on a foggy day. Everything is shrouded in white." Since emojis can also convey metaphors, we might use emojis of a mountain, mountain shoes, foggy weather, and the color white to express the same meaning. These emojis can accompany the metaphorical text as well. Another option is to send a GIF, which serves as a visual metaphor. Sending videos, audios, and other media that depict climbing a mountain on a foggy day could also be viable options.

On the other hand, video games are multimedia platforms that not only allow chatting but also enable two players to simultaneously experience the same metaphorical scene with interactive mechanisms [8, 14, 16]. If we consider that sending metaphorical texts, emojis, GIFs, and more is a way to help the listener imagine the metaphorical situation the sharer hopes to express and depict, then inviting the listener into a game scene directly reveals this imagination to them. For example, we can invite friends to climb the mountain in a foggy environment in-game and share our feelings there. However, the game still focuses on gameplay aspects : we might be disturbed by some in-game tasks or we need to set up levels before unlocking certain environments. This highlights a gap between messaging apps and video games. By crafting and utilizing *MetaphorChat* ourselves, we aim to illustrate this "in-between" space, which fuses chatting functionality with multi-modal metaphor representation (including graphics, sound, music, etc.), and interaction mechanisms for supporting emotional communication. As Kirmayer [?] describes: *"Metaphor occupies an intermediate realm, linking narrative and bodily-given experience through imaginative constructions and enactments that allow movement in sensory-affective quality space."* **Such a multimedia space is an externalization of this sensory-affective space, which has the potential to be an effective medium that amplifies this capability, allowing users to better grasp and engage with these metaphors.**

# Our approach on metaphorical chatting space design

By reflecting on the design process and incorporating feedback from user interviews, we have summarized our approach on how to integrate metaphors into a chatting space. These focus on three stages: metaphor mapping, integrating metaphors into the chat system, and tweaking the "vibe". We hope this flexible approach will inspire other researchers and practitioners.

## Step 1: Metaphor Mapping

When designing a metaphorical chatting space to communicate specific inner feelings, we might begin by considering the concept of the metaphor and identifying the source and target domains. This stage could be a purely conceptual process, not involving implementation details. Real-life experiences, poetry, films, or other literary works can serve as rich sources of inspiration, providing a diverse array of metaphors. For example, while conceptualizing **On a Train**, I drew inspiration from the real-life experience of kayaking and from playing the video game *Old Man's Journey* [34].

## Step 3: tweaking the "vibe"

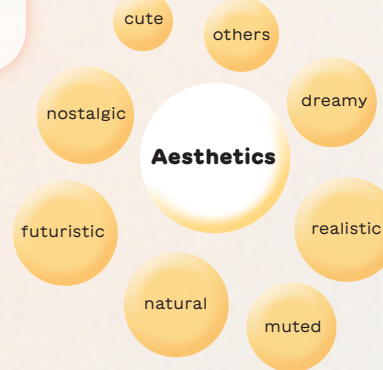

The aesthetics style of all the elements of avatar, scene background, scene object, overlaying UI, music and the sound will also influence how people perceive the chatting experience.

Maintaining *harmony of modalities* when designing digital experience [13]. *MetaphorChat* is primarily designed in a warm, cute, and dreamy aesthetic style because we hope to create a comforting space for discussing somewhat "deep" and "heavy" topics. Other styles might also evoke different feelings, which can be explored in the future.

## Step 2: Integrating metaphors into the chat system

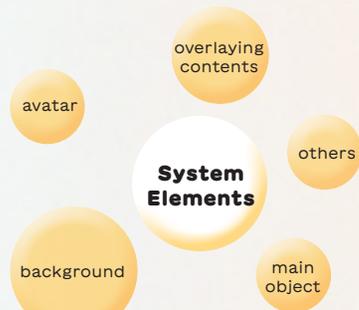

When integrating metaphor concepts into the chat system, we can try to represent the metaphors through avatars, backgrounds, main objects, or overlaying contents.

For each element, we can consider rendering them with text, graphics, sound, or music. For example, I added both graphics and train background sound to represent the background.

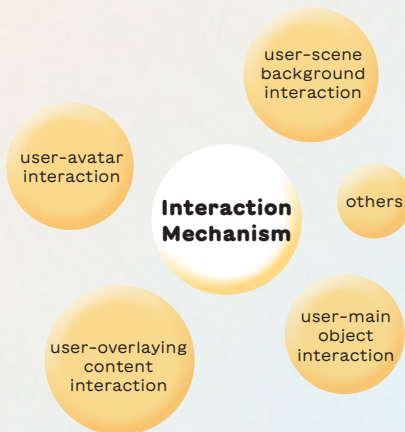

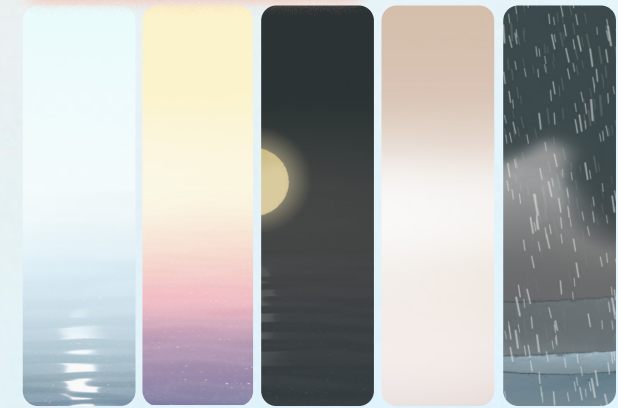

Metaphors can also be translated into interaction mechanisms. It could be the interaction between the user and avatar, scene background, main objects, overlaying contents, or other elements.

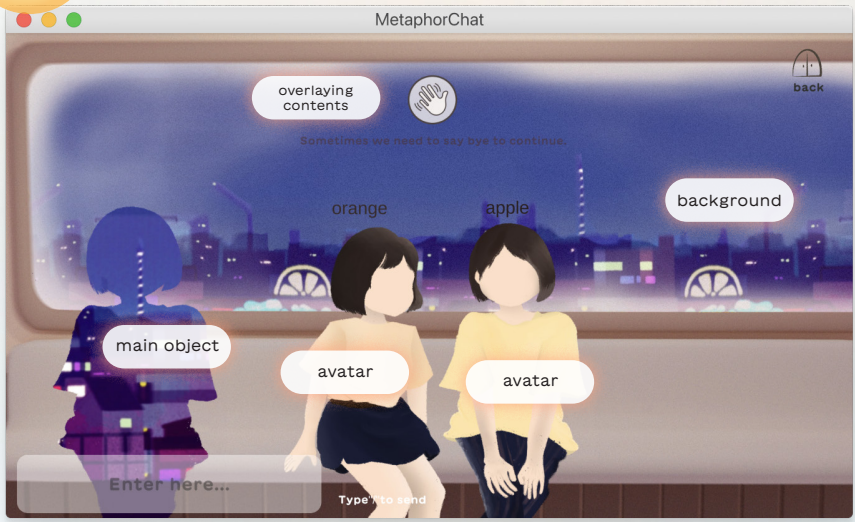

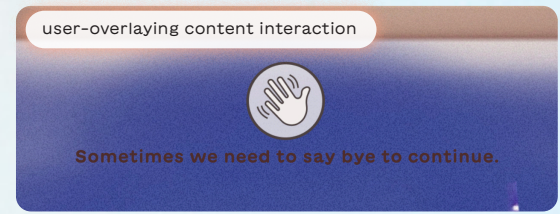

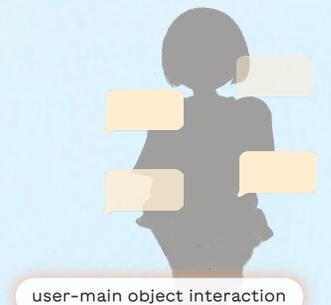

## LIMITATION AND FUTURE WORKS

While it supports the main researcher in expressing and being understood in her personal use case, the participant sample is limited—it might not apply to others due to individual variability. Nevertheless, this work, as the initial step in a broader investigation of using metaphor as a tool for empathetic and therapeutic communication design, has established a vision for future exploration. For the future generalizability of such an application, we hope to conduct several design workshops to build a collective metaphorical design library that includes concepts, graphics, sound, and interactive mechanisms that support people in expressing feelings and experiences with metaphors. Additionally, since people's interpretations are largely influenced by culture, we would also like to explore the universality and variety in designing these metaphors, such as adding customization for metaphor generation in the chatting experience.

## CONCLUSION

This project followed an autobiographical design approach to prototype *MetaphorChat*, which comprises two metaphorical chatting scenes tailored to meet researchers' genuine needs for discussing specific life topics in close relationships. The self-use of *MetaphorChat* embodies a potential slow and healing metaphorical chatting space that combines chatting functionality with graphics, sound, music, and interaction mechanisms. We envision an unexplored space between messaging apps and video games for the future design of empathetic communication applications. Our approach to designing such a metaphorical chatting space is summarized.